\documentclass{aa}
\usepackage{graphicx}
\title{Metallicity gradients in the Sagittarius dwarf spheroidal Galaxy.}
\author{C. Alard \inst{1,2}}
\offprints{C. Alard}
 \institute{Institut d'Astrophysique de Paris, 98bis Boulevard Arago, F-75014.
 \and
 Observatoire de Paris, 77 avenue Denfert Rochereau, 75014 Paris.}
\date{}
\begin{document}
\abstract{
 Metallicity gradients in the Sagittarius dwarf Galaxy (Sgr) are investigated
 by using infrared photometric data from the 2MASS survey. To search for
 metallicity effects, the giant branch in a field situated near the
 Center of the Sgr is compared to the giant branch in a field situated near
 its southern edge. The contamination of Sgr giant branch by foreground 
 Galactic stars is canceled by statistical
 subtraction of diagrams symmetrical in Galactic latitude. After subtraction
 it is possible to reconstruct the Sgr giant branch with excellent accuracy.
 The giant branch in the two fields
 have similar slopes but are shifted in color. Even after correction for 
 the differential reddening between the fields, the shift in color between
 the branch remains, and is very significant. This variation in the color
 of the giant branch corresponds to a metallicity variation of about
 -0.25 Dex. The existence of a  metallicity gradient in Sgr may indicate
 that there are two different stellar population in Sgr. One has low
 metallicity, and another one of higher metallicity has a smaller
 spatial extension.
}
\maketitle
\section{Introduction}
 The closest and largest Galaxy on the sky, the Sagittarius dwarf galaxy (Sgr)
 has been discovered only recently (Ibata, Gilmore,
 \& Irwin 1994 (IGI)). This Galaxy seems to be a dwarf spheroidal situated only 25 Kpc
 from us and only 16 Kpc from the Galactic center . Most of Sgr is situated at low Galactic latitudes
 and is seen through a dense screen of stars from the Milky Way. Even at the
 center of Sgr (b $\simeq$ -14), the density on sky of stars from the Milky
 way is overwhelming. This situation explains why the detection of
 Sgr is so difficult, and could be achieved only recently. One possible way to 
 separate stars from Sgr and the Galaxy is to measure the radial velocity.
 There is a systematic difference of about 200 kms$^{-1}$ for stars in Sgr and
 stars in the Milky way. This large difference of velocity permitted the discovery
 of Sgr by IGI. Another method to identify stars in Sgr is to search for
 RR Lyrae variables. The RR Lyrae stars are good standard candles, and
 they can be used to probed the distribution of the stellar density as a function
 of distance. The histogram of the RR Lyrae magnitude of a field  in Sgr shows a
 double peaked distribution. The first peak corresponds to the Galaxy, while
 the second peak corresponds to Sgr (Alard 1995). By selecting RR Lyrae in 
 the second peak it is possible to map the spatial distribution of Sgr.
 Using this method it was possible to identify a new extension of Sgr at lower
 Galactic latitudes (Alard 1996). This work was recently extended in order
 to produce a large map of Sgr, from the lower latitude extension to the center
 of the Galaxy (Alard, Cseresnjes,\& Guibert 2000). In addition to the radial
 velocities and the RR Lyrae, there are photometric methods that can be used
 to probe the structure of Sgr. For instance, IGI used a feature in a $B_J$ vs.
 $B_J$-R color magnitude diagram to identify stars in Sgr. However the
 method is not very efficient at lower Galactic latitude, due to contamination
 by the Galaxy. Using similar color magnitude diagrams, Bellazzini,
 Ferraro, \& Buonanno (1999) were able to further the study of Sgr
 stellar populations. They demonstrate that a very metal-poor population
 is present in Sgr, and they show possible hints for a metallicity 
 gradient in Sgr. The advent of new infrared data from the 2MASS survey
 offers is very promising for the study of Galactic structure in
 general. There are well-defined structures in the infrared color
 magnitude diagrams, like the upper giant branch, which can
 be very useful to probe the structure of Galaxies. Furthermore,
 the dramatic weakening of the interstellar extinction obtained
 by going to the infrared is a very important asset. We will see
 that these infrared color magnitude diagrams are a efficient
 tool used to study the stellar population of Sgr. 
\section{The Data}
 Two fields have been selected to investigate the metallicity effects in Sgr.
 One of the fields is close to the center of the Galaxy, while the other
 is close the lower latitude edge of the Galaxy according to the map
 of Ibata {\it et al.} (1995) (see Fig. 1).
 All the stellar magnitudes have been extracted from the 2MASS point 
 source catalogue. The stellar populations in these fields are well
 characterized by their K vs. (J-K) color magnitude diagrams . 
 To identify features or sequences which are not related to the Galactic
 population, it is interesting to compare these diagrams with diagrams
 obtained from fields symmetrical in Galactic coordinates (see table 1
 for a summary of the fields location). By comparing
 the symmetrical diagrams we see immediately that for the fields in Sgr,
 an additional sequence is present at the right of the diagrams (see Fig 2). 
 This sequence is visible for the field at the center of Sgr, but also 
 for the field near the lower latitude edge. This feature is certainly
 associated with the upper giant branch of Sgr. The tip of the sequence
 is approximately at K$=$10.5 which is consistent with an object at
 a distance of 25 Kpc. 
\begin{table} 
   \caption{The fields}
   \label{}
   \begin{tabular}{lll}
      \hline\noalign{\smallskip}
   Field & field center & size of field\\ 
   \hline\noalign{\smallskip}
     Center & (l,b)$=$(6,-14) & ($\Delta$l,$\Delta$b)$=$(2,2) \\ 
     Edge & (l,b)$=$(6.5,-20) & ($\Delta$l,$\Delta$b)$=$(2,2) \\ 
     Center-sym & (l,b)$=$(6,14) & ($\Delta$l,$\Delta$b)$=$(2,2) \\  
     Edge-sym & (l,b)$=$(6.5,20) & ($\Delta$l,$\Delta$b)$=$(2,2) \\  
   \hline\noalign{\smallskip}
      \noalign{\smallskip} 
  \end{tabular}
\end{table}
\begin{figure*} 
\centering
\includegraphics[width=15cm]{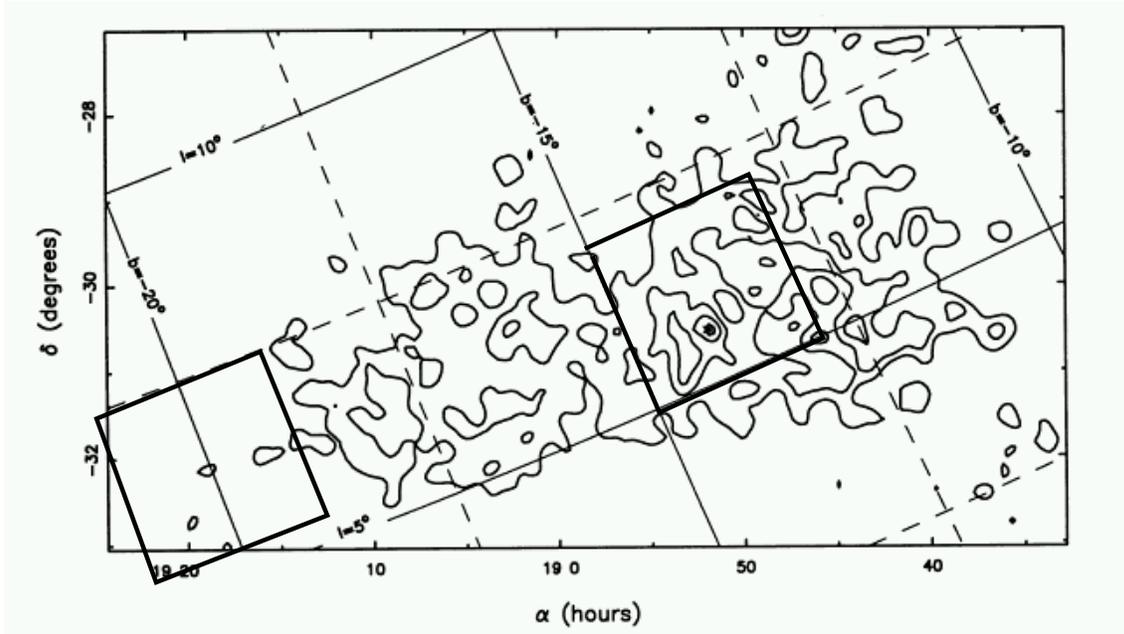}
\caption{The position of the 2 fields in Sgr, superimposed on
 the map of Ibata {\it et al.} (1995)}
\end{figure*}
\begin{figure*} 
\centering
\includegraphics[width=15cm,angle=-90]{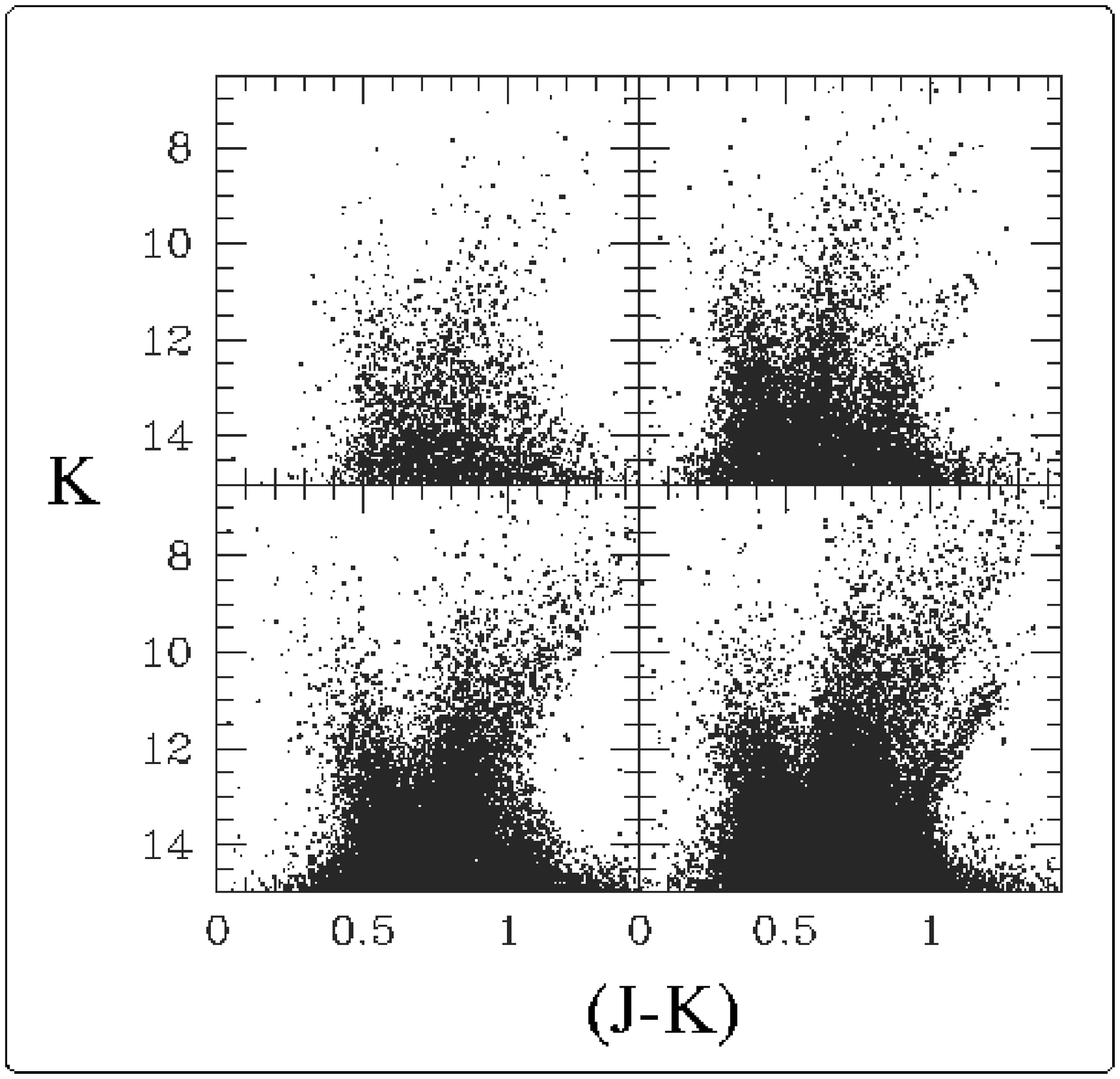}
\caption{The K vs. (J-K) color magnitude diagrams for the 4 fields.
 The fields in Sgr are on the right side of the diagram,
 while the field at the symmetrical position in Galactic coordinates
 are to the left. The upper right diagram is for the field near
 the edge of Sgr, and the lower right diagram for a field
 at the center of Sgr.}
\end{figure*}
\section{Reconstruction of Sgr giant branch}
 The Sgr giant branch is nicely visible in the color magnitude diagrams
 of our two fields. In general, the contamination of the Sgr giant
 branch by stars from the Galaxy is a serious problem. However,
 for bright stars (K$>$12) there is very little contamination of the Sgr
 giant branch by foreground Galactic stars. Thus it is possible
 to use this bright tip of the giant branch to estimate its width. 
 Before looking at the width of the branch, it is interesting to apply
 a small rotation to the diagram in order to have the branch vertical.
 This can be achieved easily by defining the new color index: 
$(J-K)_E = (J-K)-\frac{K-11}{50} $
 .\ \ By making an histogram using the modified color index $(J-K)_E$ 
 for stars brighter than (K$=$12), and by fitting a gaussian
 around the position of the branch, we can estimate the width
 of the branch. The results of the fitting are summarized in Table 2.
 There is no significant difference in the width of the branch between
 the two fields. This width is somewhat larger than the internal scatter,
 as derived from the errors on measuring the magnitudes, quoted in this
 release of the 2MASS catalogue. It shows that the mean scatter in metallicity
 in Sgr is probably quite small.
\begin{table} 
   \caption{The two fields}
   \label{}
   \begin{tabular}{lll}
      \hline\noalign{\smallskip}
   {\bf Field} & {\bf Position of the branch} & {\bf Width} \\ 
   \hline\noalign{\smallskip}
     Sgr center & $(J-K)_E=1.07$ & $\sigma=0.0484$  \\
     Sgr edge & $(J-K)_E=0.99$ & $\sigma=0.0498$ \\
   \hline\noalign{\smallskip}
      \noalign{\smallskip} 
  \end{tabular}
\end{table}
 To proceed further in our analysis, we need to estimate the slope
 and the position of the giant branch in the two fields. 
 It is not possible to restrict our analysis to the brighter
 stars to estimate the shape of the branch. The dynamical range would not be 
 sufficient. Thus we need to extend our fitting procedure to the fainter
 stars which are contaminated by the Galactic foreground stars.
  A simple solution to this contamination problem is
 to subtract the contribution from the Galaxy by assuming that the Galaxy is 
 symmetrical in latitude. 
\subsection{Subtraction of the diagrams.}
 In practice, the subtraction can be performed by binning the data. We use
 the same bin size for all the data in K; the size of the bin is 0.1 mag. 
 and is 0.025 mag in (J-K). 
 Before we can subtract the diagram we have to  
 compensate for the differential reddening between the fields. 
 This differential reddening can be estimated by searching for the maximum
 correlation between parts of the diagrams which are far from the giant
 branch. The area we selected for cross correlation is presented in Fig. 3. 
 Once the reddening alignment has been performed, we subtract the diagrams,
 and normalize the residuals by the square root of variance of the noise
 in the subtracted image. The noise is estimated in the region of the diagram
 that we already used to estimate the differential reddening. As we expect
 from Poisson statistics, the variance of the noise is correlated to the
 counts in the initials diagrams. However the statistics of the
 residual is larger than the Poissonian expectation the reason for this
 is probably that the reddening variation as a function
 of distance is nor perfectly identical between symmetrical Galactic fields.
 Another possibility is that the distribution of stars in the Galaxy 
 is not perfectly symmetrical. It is possible that the Galactic bar
 is slightly tilted out of the plane, or that the structure of the
 spiral arms are not perfectly smooth. These small asymmetries are sufficient
 to create an additional source of noise in the diagrams. 
  However, it is important to note that 
 these additional fluctuations are unlikely
  to create significant systematic biases. First, there are no reason
 that the (small) effect of extinction be correlated between the two fields.
 Secondly, even if some asymmetries exist in the Galaxy between
 positive and negative longitude, the effect cannot be large (the amplitude
 of a spiral arms is only 5 to 10 \% of the density). Even in the low density
 regions near the edge of Sgr, the density of Sgr giant branch in the
 CMD diagram is comparable to the density of Galactic stars, thus the 
systematic effects should stay beyond 10 \%, which is not much larger than the
 basic statistical fluctuations (Poisson noise).
 Finally, 
 the subtracted diagrams normalized by the noise expectation are smoothed
 using a 3$\times$3 mesh. The two diagrams we obtain are presented in Fig. 4.
\subsection{Location of the giant branch.}
Once the diagrams have been decontaminated by the subtraction
 process, we can measure with good accuracy the position of the Sgr 
 giant branch. Since the branch is almost vertical, its location
 can be found by estimating the position of its intersection 
 with a horizontal line. In the K vs. (J-K) plane this is almost equivalent
 to searching for the maximum of the histogram in (J-K) of a
 strip defined by the condition: $K_0<K<K_0+\Delta K$. 
  This histogram corresponds to a line in the image of the cmd presented in Fig. 4 
 (reconstructed by binning the cmd). An estimate 
 of the giant branch position
 is given by the position of the pixel with maximum counts. Let us call
 $nx_1$ the value of the pixel at maximum, and $x_1$
 the position of this pixel (integer number).  This position
 has an accuracy which is not better than the pixel size. It is possible to improve the accuracy
 by using the neighboring pixels, with positions
 ($x_0=x1-1$, $x_2=x1+1$), and associated pixel values ($nx_1$, $nx_2$). 
 To refine the position, we will use a parabolic interpolation.
 Then the location of the maximum can be calculated by adding 
 a fraction of a pixel shift given by the formula:
 $\delta_X=0.5 \frac{(nx_0-nx_2)}{nx_2+nx_0-2 \ nx_1}$.
 The position of the giant branch in the two field has been
 reconstructed using this method (Fig. 5). The slope of giant branch
 in the two fields is not significantly different.
  However, it is clear that the giant branch 
 corresponding to the field near the edge of Sgr is redder than at the center
 of Sgr. However, before we can claim any difference in the color of the
 giant branch between these two fields, we have to compensate for the 
 differential reddening between the center and the edge of Sgr.
\begin{figure} 
\centering
\includegraphics[width=9cm]{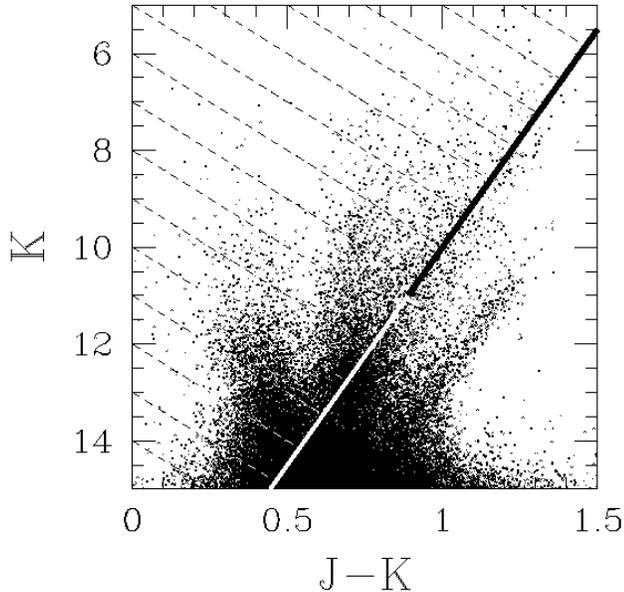}
\caption{Color magnitude diagram of the field near
 the center of Sgr. The area of the diagram we used to estimate 
 the differential reddening between the fields is indicated by dashed
 lines.}
\end{figure}
\begin{figure*} 
\centering
\includegraphics[width=13cm]{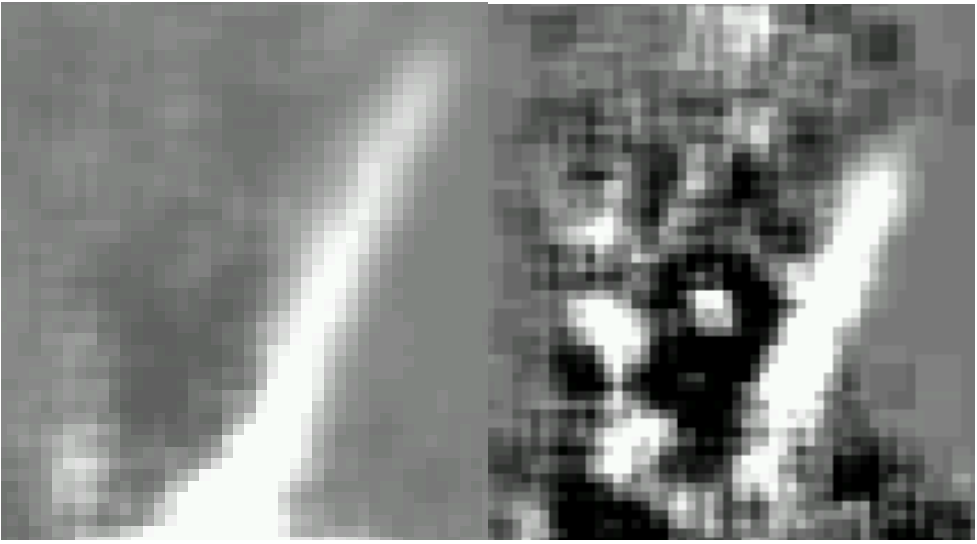}
\caption{The subtracted images obtained by using the procedure
 described in Spec. 3.1. Left is the subtracted image corresponding
 to the field near the center of Sgr. The other image corresponds to the
 field near the edge of Sgr, near b$=$-20.}
\end{figure*}
\begin{figure} 
\centering
\includegraphics[width=9cm]{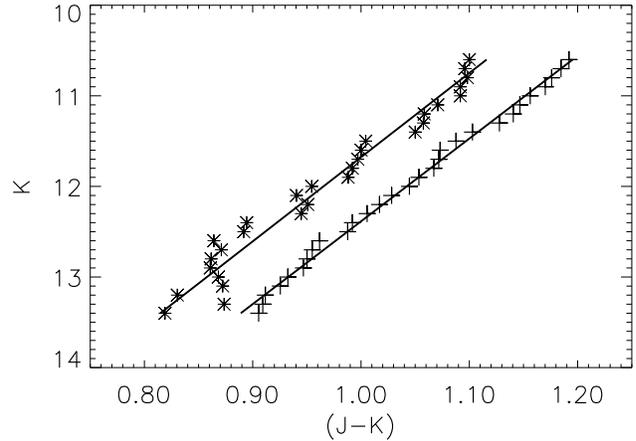}
\caption{Reconstruction of the giant branch in the two fields. The location
 of the giant branch as been estimated by using the subtracted diagrams
 presented in Fig. 4.}
\end{figure}
\subsection{Differential reddening between the 2 fields in Sgr.}
 To estimate the differential reddening between the field at the center
 of Sgr and the field near the edge, we will use the stars situated
 in the upper left region of the color magnitude diagram. The upper
 left region will be defined by the following conditions: \linebreak
 (J-K)-(K-11)$/50<0.8$ and K$<12$. \
The upper left region of the color magnitude diagrams is occupied
 by main sequence stars located in foreground regions of the Galactic disk. 
 These stars are relatively close (a few Kpc), but at latitude
 $b>14^{\circ}$ the line of sight escapes very quickly
 the thin layer where the interstellar material is concentrated.
 Thus it is very unlikely that reddening occur beyond a few
 Kpc from the sun for our two fields, and consequently, the differential 
 reddening of Sgr can be estimated by using foreground disk stars
 situated a few Kpc from us. One may also wonder
 metallicity gradients in the disk
 effects can bias our differential estimation of the reddening.
 However, between the field near Sgr center and
 the field at the edge of Sgr, the difference in height above the plane
 is less than 0.1 Kpc. It is unlikely that such small difference
 in height above the plane will result in a systematic difference
 in color (due to metallicity effects) between the two lines of sight.
 A metallicity gradient in Sgr cannot be mimicked by a metallicity
 gradient in the Galaxy. The effects have opposite directions.
 Even if a small difference in metallicity existed, it would
 result only in a slight under-estimation of the metallicity gradient 
 between the two fields in Sgr. \linebreak
 The differential reddening is estimated by cross correlation between 
 the sequences of foreground stars in the two fields. 
 The cross correlation is estimated after binning the data along the color
  axis. The color has been slightly modified to take into account the
 slope of the sequence using our former
 relation (Color$=$(J-K)-(K-11)$/50<0.8$). The maximum correlation
 as a function of the differential reddening between the fields is estimated
 by fitting a parabola to the data (see Fig. 6). The estimated differential
 reddening between the fields is 0.031 mag with an error close to 0.0015 mag.
 Considering that the uncorrected difference in color between the two
 giant branches is 0.075 mag, the reddening$-$corrected difference in color is
 0.044 mag. The total error on this differential color is only about 0.006 mag.
 There is no doubt that we have found a very significant color effect
 in Sgr. 
  However, before claiming a systematic color difference between the two giant 
branches, one may wonder if this shift cannot be interpreted as a 
 differential distance effect between the two fields. A color shift
 of 0.044 mag in (J-K) corresponds to a shift in K of about $0.044 \times $
 (GB Slope) $\simeq$ 0.45 mag. At the distance of Sgr it corresponds to
 a difference in distance of about 5 Kpc. However, the projected distance
 between the field is only of $\simeq$ 2.5 Kpc. The projected distance
 is an upper limit, in reality, due to the extension of the Sgr body, even 
if Sgr is highly inclined, the mean effect will be smaller.
 Thus this shift in color cannot be interpreted as an effect of distance.
 Note also that this systematic shift is statistical. This is a mean
 difference between a large number of stars. Thus even if some spread
 in distance  exists between individual stars due
 to the extent of Sgr along the line of sight (as illustrated by Bonifacio
 {\it el al.} 2000, Ibata {\it et al.} 1997, and Helmi \& White(2001)), what
 we measure, which is the mean location of the giant$-$branch 
 is well defined statistically.
 The internal scatter in distances within Sgr will result in some
 broadening of the giant branch. According to the result given
 in Table 2, an internal scatter of a few Kpc within Sgr body is possible,
 but it is very hard to be more specific, since we are limited by
 the photometric errors.

\begin{figure} 
\centering
\includegraphics[width=9cm]{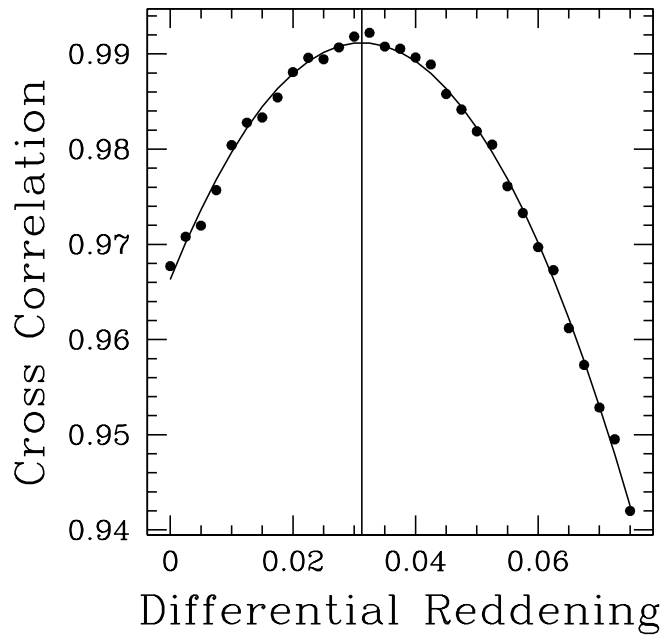}
\caption{Cross correlation between the sequence of foreground stars
 for the two fields in Sgr. The cross correlation is plotted as a function
 of the differential reddening between the field near the edge and the field
 near the center of Sgr.}
\end{figure}
\section{Measuring the metallicity gradient in Sgr.}
 We detected a significant shift in the color of the giant branch 
 between the center and edge of Sgr. This shift can be due either
 to a systematic variation in age or metallicity of the stellar
 population in Sgr.
 One may interpret this color shift as a metallicity variation
. The metallicity inferred
 from the slope of the giant branch using the Kuchinski \& Frogel (1995) relation are:
 $[Fe/H]=-2.98-23.84  {\rm \times (GB \ Slope)}=-0.37 \pm 0.04$  near Sgr center, and $[Fe/H]=-0.41 \pm 0.1$
 at the edge of Sgr (note the errors quoted here are internal errors). 
 First, we note that the metallicity inferred using the Kuchinski relation for Sgr
 giants is not far from the metallicity measured by Bonifacio for two
 giants in Sgr ( $[Fe/H]=-0.28$ and $[Fe/H]=-0.21$ ). This metallicity
 estimation is also consistent with the metallicity calculated by Dudziak
 {\it et al.} (2000) who found $[Fe/H]=-0.55$ 
 Some other studies seems to reveal a mixture of populations
 in Sgr. For instance, Smecker-hane, Mc William \& Ibata (1998) 
 analyzed the spectra of 7 stars in Sgr. Some of the stars show
 solar metallicity abudances, while some have metallicity
 comparable to stars in the Galactic Halo. Bellazzini, Ferraro, \& Buanono
 (1999) found similar results by analyzing V, (V$-$I) color magnitude
 diagrams in Sgr. Finally, a metal poor population was identified
 in M54 (Brown, Wallerstein, \& Gonzalez 1999).
 Thus, measuring a metallicity in Sgr may depend on the
 type of the stellar population under investigation. The result we obtained
 indicates that our sample of giants belongs to the metal rich population 
 in Sgr, but obviously we do not detect any significant 
 metallicity variations between the two fields by this method. 
 However, in Sec. 3.3 we found a very significant difference in the color
 of the branch in the two fields. According to the linear relation presented 
 in Fig. 8, this color shift corresponds to about 0.2 Dex in metallicity. 
 This result is consistent (within the uncertainties in measuring the slope)
 with  our previous measurements of the
 metallicity using the Kuchinski \& Frogel (1995) relation.
 We can conclude that mean metallicity of our Sgr giants 
 seems to be about -0.4 Dex, with a systematic trend of about 0.2 Dex
 from the center to the edge of the Galaxy. Concerning internal 
 metallicity dispersion within each field, we remind
 the reader that the width of the giant branch is about 0.05 mag in the (J-K)
 color index (Sec. 3). According to the linear relation presented in Fig. 8,
 this
 width corresponds to about 0.25 Dex in metallicity. We may conclude that
 the internal dispersion in metallicity is about 0.25 Dex. However, 
 it is possible that a small fraction of the total number of stars
 has a much lower metallicity that the mean. Such a small tail in the
 metallicity distribution is very hard to identify using color magnitude
 diagrams.
\section{Discussion}
 We may also interpret our results as a possible age variation. However,
 the effect of age on color is weak.
 For instance, at a metallicity
 close to -0.4 Dex, using the theoretical isochrones of
 Bertelli {\it et al.} 1994, \& Girardi {\it et al.} 1996 we find that 
 the maximum difference in color obtained for 17.4 Gyr
 and 6.8 Gyr is 0.056 mag in (J-K) (see Fig. 7). Thus, it is possible that
 age has an effect on the observed color shift, but this effect
 is probably not dominant. Thus we can conclude that the stellar
 population in Sgr is slightly older and slightly more metal
 poor near the edge of Sgr than at its center, with a metallicity
 effect close to 0.2 Dex. 
  It is important to notice that similar
 metallicity gradients have been found in other dSph. Da Costa
 {\it et al.} 1996 showed by analyzing HST color$-$magnitude
 data from satellites of
 M31 that the morphology of the horizontal branch (HB) differs
 from the center to the outer parts for 2 out of 3 of the dSph's 
 in the sample. This change in the HB morphology suggests the
 possibility of a metallicity gradient. Similarly, Hurley$-$Keller,
  Mateo \& Grebel (1999) describe a change in the HB morphology in their
 study of the Sculptor dwarf galaxy.
 One may interpret the metallicity 
 gradient in Sgr as a smooth variation of the chemical composition of
 a single stellar population in Sgr. This might be more
 easily represented if we assume that Sgr is composed of two
 different stellar populations. These two populations could 
 be a metal$-$poor component (Halo), and a more metal$-$rich
 component with less spatial extension. The combination
 of both population is such that the Halo becomes dominant
 only near the edge. One may wonder what the more
 metal rich component could be. If we compare this Galaxy to a dwarf
 Galaxy with similar metallicity (The LMC) we find that
 this more metal$-$rich component may look like a disk. Since disks
 are fragile to dynamical perturbations, in the
 case of Sgr we would observe only the remains of a tidally disrupted
 disk. 
  This scenario is consistent with the dynamical simulations recently
 performed by Mayer {\it et al.} 2001 which show that small
 irregular galaxies constituted of a disk and a halo of dark matter
 are transformed into dSph by tidal processes after only 2$-$3 orbits
 around the Galaxy.
  This discussion also
  raises the problem of the intrinsic nature of the Sgr dSph galaxy. While
  it currently is thought to be a dwarf spheroidal system, the presence of a pronounced
  radial gradient in metallicity suggests the possibility that before
 its tidal disruption Sgr could have been of a different type.
%
%
%
%
%
%
\begin{figure} 
\centering
\includegraphics[width=7cm]{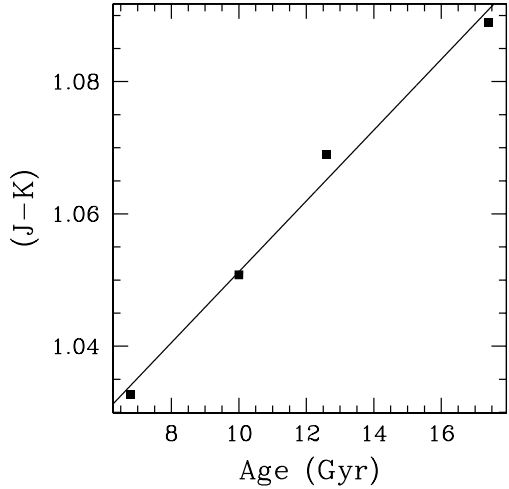}
\caption{The color of the upper giant branch near K$=$12 ($\rm K_0=-2.5$),
 estimated from the theoretical isochrones of Bertelli {\it et al.} 1994, \& Girardi {\it et al.} 1996.
 We represent the color of the giant branch as a function of age
 for a metallicity $[Fe/H]=-0.37$. The color vs. age variation was approximated
 by a linear relation.}
\end{figure}
\begin{figure} 
\centering
\includegraphics[width=9cm]{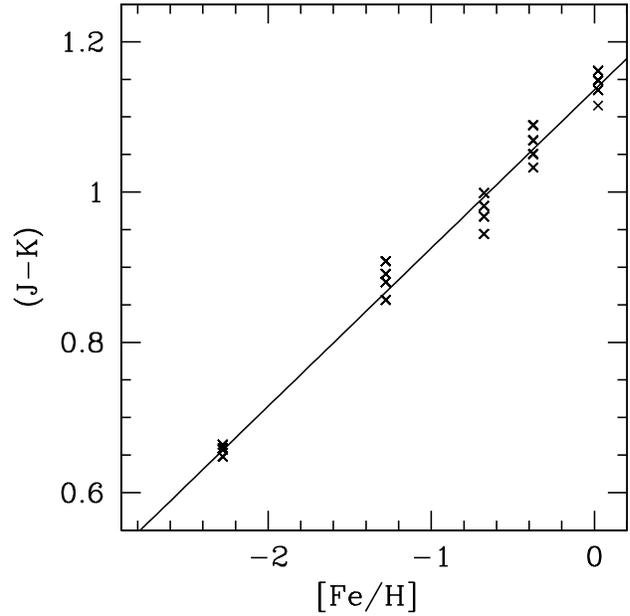}
\caption{The color of the upper giant branch near K$=$12 ($\rm K_0=-2.5$) 
 estimated from the theoretical isochrones of
  Bertelli {\it et al.} 1994, \& Girardi {\it et al.} 1996.
 We represent the color of the giant branch as a function of metallicity
 for four different ages (6.3, 10, 12.6, 17.4 Gyr). Note that the color
 vs. metallicity variation is well approximated by a straight line.
}
\end{figure}
\begin{acknowledgements}
I am pleased to thank Jay Gallagher for interesting suggestions.
\end{acknowledgements}

\label{lastpage}

\end{document}